&pdflatex

\documentclass[12pt,prd,aps,amssymb,amsmath,tightenlines,showpacs]{revtex4}
\usepackage{graphicx,epsfig,color}
\usepackage{comment}

\def\cPT{\mathcal PT}
\def\cO{\mathcal O}

\graphicspath{{FIGURE/}}
\begin{document}

\title{$\cPT$-symmetric $\varphi^4$ theory in $d=0$ dimensions}

\author{Carl M. Bender$^a$}\email{cmb@wustl.edu}
\author{Vincenzo Branchina$^b$}\email{branchina@ct.infn.it}
\author{Emanuele Messina$^b$}\email{emanuele.messina@ct.infn.it}

\affiliation{${}^a$Department of Physics, Washington University, St. Louis, MO
63130, USA\\
${}^b$Department of Physics, University of Catania and INFN, Sezione di
Catania, Via Santa Sofia 64, I-95123 Catania, Italy}

\begin{abstract}
A detailed study of a $\cPT$-symmetric zero-dimensional quartic theory is 
presented and a comparison between the properties of this theory and those of a
conventional quartic theory is given. It is shown that the $\cPT$-symmetric
quartic theory evades the consequences of the Mermin-Wagner-Coleman theorem
regarding the absence of symmetry breaking in $d<2$ dimensions. Furthermore, the
$\cPT$-symmetric theory does not satisfy the usual Bogoliubov limit for the
construction of the Green's functions because one obtains different results for
the $h\to0^-$ and the $h\to0^+$ limits.
\end{abstract}

\date{\today}
\maketitle
\section{Introduction}
\label{s1}
This paper examines the structure of $\cPT$-symmetric \cite{r1} $-\varphi^4$
theory in $d=0$ dimensions and makes detailed comparisons between this theory
and conventional $\varphi^4$ theory. This work is a natural sequel of our
studies of $\cPT$-symmetric $i\varphi^3$ theory \cite{r2,r3}. We will see that
the structure of the $-\varphi^4$ theory is much richer and more elaborate than
that of conventional $\varphi^4$ theory and also of the cubic theories.

This paper is organized as follows. In Sec.~\ref{s2} we review the main features
of the conventional $\varphi^4$ theory in $d=0$ dimensions in the presence of an
external linear source $h$. The partition function for this theory is $\int\!d
\varphi\,e^{-V(\varphi)}$, where $V(\varphi)=h\varphi+m^2\varphi^2/2+g\varphi^4
/24$. We analyze this theory using the method of steepest descents. There is a
critical value of $m^2$: For $m^2>m^2_c$ all three saddle points are real, while
for $m^2<m^2_c$ one saddle point is real and the other two form a
complex-conjugate pair. In both cases semiclassical approximations to the
partition function and the corresponding Green's functions are obtained. Because
$d<2$, the one-point Green's function vanishes even when $m^2<m^2_c$. (When $d
\geq2$, if $m^2<m^2_c$ symmetry breaking occurs.)

In Sec.~\ref{s3} the corresponding analysis is performed for the
$\cPT$-symmetric theory in which $g$ and $h$ are replaced by $-g$ and $ih$.
Again, we find a critical value of $m^2$. For $m^2>m^2_c$, one of the three
saddle points is purely imaginary and the other two are $\cPT$-conjugate. For
$m^2<m^2_c$ all three saddle points lie on the imaginary-$\varphi$ axis. When
$m^2>m^2_c$, new and surprising features appear. We find that for $h>0$ only one
saddle point contributes to $Z[h]$, while for $h<0$ all three saddle points
contribute. As a consequence, the partition function $Z[h]$ for $h>0$ differs
from the partition function for $h<0$ and the Green's functions obtained by
taking the limit $h\to0^-$ differ from those obtained in the limit $h\to0^+$.
Thus, the $\cPT$-symmetric quartic theory evades the properties of the
Bogoliubov limit for the construction of the Green's functions. (The Bogoliubov
calculation consists of taking derivatives of the vacuum persistence amplitude
with respect to the external source $h$, and then performing the $h\to0$ limit,
irrespective of the sign of $h$.)

When $m^2<m^2_c$, the Green's functions obtained in the $h\to0^+$ and $h\to0^-$
limits are the same. However, unlike the conventional quartic theory, the
one-point Green's function $G_1$ is nonvanishing even though $d=0$; the Green's
function has a purely imaginary value. The nonvanishing of $G_1$ contradicts the
Mermin-Wagner-Coleman (MWC) theorem on the absence of symmetry breaking in $d<2$
dimensions \cite{r9}. We also find that when $m^2>m^2_c$, the one-point Green's
function is nonvanishing. Section~\ref{s4} gives a summary and conclusions.

\section{Conventional $\varphi^4$ theory}
\label{s2}
In this section we review the ordinary $\varphi^4$ theory in $d=0$ dimensions so
that we can compare it later with the corresponding $\cPT$-symmetric theory. We
begin by considering the generating functional for the Green's functions, which
is given by the (normalized) integral 
\begin{eqnarray}
Z[h]=\sqrt{\frac{|m^2|}{2\pi}}\int^{+\infty}_{-\infty}d\varphi\,e^{-V(\varphi)}.
\label{e1}
\end{eqnarray}
The potential is given by
\begin{equation}
\label{e2}
V(\varphi)=h\varphi+\frac{m^2}{2}\varphi^2+\frac{g}{24}\varphi^4, 
\end{equation}
where $g$ is assumed to be positive to guarantee the convergence of (\ref{e1}),
but both $m^2$ and $h$ can take positive and negative values. The normalization
coefficient $\sqrt{|m^2|/(2\pi)}$ in (\ref{e1}) is chosen so that for the
noninteracting theory
$$Z_{free}[0]=\sqrt{\frac{|m^2|}{2\pi}}\int^{+\infty}_{-\infty}d\varphi\,e^{-
\frac{m^2}{2}\varphi^2}=1.$$
There is an implicit factor of $1/\hbar$ in the exponent in the integral
(\ref{e1}). In this paper we calculate the partition-function integral in the
semiclassical approximation for which $\hbar$ is small. To do so we use the
method of steepest descents \cite{r4}. We then identify the Green's functions as
coefficients of powers of $h$ in the expansion of the partition function $Z[h]$.

In order to evaluate the integral in (\ref{e1}) asymptotically in the limit
$\hbar\to0$, we consider the equation that determines the location of the
saddle points:
\begin{equation}
V'(\varphi)=h+m^2\varphi+\frac{g}{6}\varphi^3=0.
\label{e3}
\end{equation}
When solving (\ref{e3}), we note that two different situations arise depending
on the value of $m^2$. Specifically, by defining the critical value $m^2_c=-
\left(9gh^2\right)^{1/3}/2$, we see that the qualitative behavior of the three
solutions of (\ref{e3}) depends on $m^2$ being larger or smaller than $m^2_c$.
We begin by considering the case $m^2>m^2_c$.

\subsection{The $m^2>m^2_c$ case}
For $m^2>m^2_c$ there is one real solution $\varphi_0$ and there are two
complex-conjugate solutions $\varphi_+$ and $\varphi_-=\varphi_+^*$ to
(\ref{e3}). For small $h$ these solutions are
$$\varphi_0=-\frac{h}{m^2}+\cO(h^3),\quad \varphi_\pm=\frac{h}{2m^2}\pm i\left(
\sqrt{\frac{6 m^2}{g}}+\frac{h^2}{8}\sqrt{\frac{3g}{2m^{10}}}\right)+\cO(h^3).$$
For $h>0$ ($h<0$) the solutions form a triangle with one vertex on the negative
(positive) side of the real-$\varphi$ axis and the other two vertices in the
half-plane ${\rm Re}\,\varphi>0$ (${\rm Re}\,\varphi<0$).

The integral in (\ref{e1}) is approximated by using Laplace's method in the
limit $\hbar\to0$. On the real axis, there is only one {\it Laplace point} [zero
of $V'(\varphi)$], namely $\varphi_0$, which is a minimum of the potential
$V(\varphi)$. Thus, for all $h$ the partition function $Z[h]$ is approximated as
\begin{equation}
\label{e4}
Z[h]=\exp\left[-V(\varphi_0)-\frac{1}{2}\ln\left(V''(\varphi_0)/m^2\right)
\right].
\end{equation}
Inserting $\varphi_0$ into (\ref{e4}) and expanding in powers of $h$, we get
the formal Taylor series 
\begin{equation}
\label{e5}
Z[h]=1+\left(\frac{1}{m^2}-\frac{g}{2m^6}\right)\frac{h^2}{2}+\cO(h^4).
\end{equation}

As a consequence of the normalization [see (\ref{e1})] the coefficients of
powers of $h$ in (\ref{e5}) give the Green's functions of the theory. The first
two connected Green's functions $G_1^c=G_1$ and $G_2^c=G_2-G_1^2$ are
$$G_1^c=0,\quad G_2^c=\frac{1}{m^2}-\frac{g}{2m^6}.$$
These expressions are the usual perturbative result (in powers of the coupling
constant) \cite{r5}.

\subsection{The $m^2<m^2_c$ case}
For $m^2=-\mu^2<m^2_c$ all solutions to (\ref{e3}) are real:
\begin{equation}
\varphi_0=\rho\cos\left(\frac{|\theta|-2\pi }{3}\right),\quad
\varphi_\pm=\rho\cos\left(\frac{|\theta|+2\pi \,\Theta(\pm h)}{3}\right),
\label{e6}
\end{equation}
where $\Theta(h)$ is the Heaviside step function
\begin{equation}
\Theta(h)=\left\{\begin{array}{cc} 1&(h>0),\\ \frac{1}{2}&(h=0),\\ 0&(h<0),
\end{array}\right.\nonumber
\end{equation}
and
$$\rho=-2\,{\rm sgn}(h)\sqrt{\frac{2\mu^2}{g}},\quad \theta=-{\rm sgn}(h)\,
{\rm arctan}\left(\sqrt{-1+ \frac{8\mu^6}{9gh^2}}\right).$$
Note that $\varphi_-$ always lies on the negative-$\varphi$ axis while
$\varphi_+$ always lies on the positive-$\varphi$ axis. For small values of $h$
in (\ref{e6}) we find that
$$\varphi_0=\frac{h}{\mu^2}+\cO(h^3),\quad\varphi_\pm=\pm\sqrt{\frac{6\mu^2}{g}}
-\frac{h}{2\mu^2}\mp\sqrt{\frac{3g}{2\mu^{10}}}\,\frac{h^2}{8}+\cO(h^3).$$

By examining the second derivative of the potential $V''(\varphi)=m^2+g
\varphi^2/2$ at these three saddle points, we find that on the real-$\varphi$
axis, $\varphi_\pm$ are two local minima and that $\varphi_0$ is a local maximum
of $V$. For $h>0$ the absolute minimum is at $\varphi_-$ while for $h<0$ the
absolute minimum is at $\varphi_+$. Consequently, the contribution to the 
asymptotic behavior of $Z[h]$ in (\ref{e1}) from $\varphi_\pm$ is given by
$$\frac{Z^{(\pm)}[h]}{Z[0]}=\exp\left[-V(\varphi_\pm)-\frac{1}{2}
\ln\left(V''(\varphi_\pm)/\mu^2\right)\right].$$
Observe that when $h>0$ ($h<0$) the asymptotic behavior of $Z[h]$ is dominated
by $\varphi_-$ ($\varphi_+$) and that the contribution from $\varphi_+$
($\varphi_-$) is subdominant (exponentially small). Expanding the above
formula in powers of $h$, we obtain
\begin{eqnarray}
\frac{Z[h]}{Z[0]}&=& \left(1+\delta_{h,0}\right) \Bigg[1
+\left(\Theta(-h)\sqrt{\frac{6\mu^2}{g}} -\Theta(h)\sqrt{\frac{6\mu^2}{g}}
+\Theta(h)\frac{1}{4}\sqrt{\frac{3 g}{2\mu^6}}
-\Theta(-h)\frac{1}{4}\sqrt{\frac{3 g}{2\mu^6}}\right)h\nonumber\\
&+&\left(-\frac{1}{\mu^2}+\frac{6\mu^2}{g}
+\frac{13}{32}\frac{g}{\mu^6}\right)\frac{h^2}{2}+\cO(h^3)\Bigg].
\label{e7}
\end{eqnarray}
This expression is discontinuous at $h=0$. A doubling occurs at $h=0$ because at
this special point there are no subdominant contributions; that is,
$\varphi_\pm$ contribute equally to the semiclassical result (\ref{e7}) when
$h=0$.

Finally, we identify the connected Green's functions $G_1^c$ and $G_2^c$
from (\ref{e7}):
$$G_1^c=0,\quad G_2^c=-\frac{1}{\mu^2}+\frac{6\mu^2}{g}+\frac{13g}{32\mu^6}.$$
Note that for $d=0$ dimensions, the one-point Green's function $G_1^c$ vanishes
even when $m^2<m^2_c$. As is well known, when $d<2$ there is no spontaneous
symmetry breaking because there is tunneling between the two minima $\varphi_+$
and $\varphi_-$ of the potential \cite{r6}. In our case ($d=0$) the vanishing of
$G_1^c$ is due to the cancellation between the two contributions to $Z[h]$
coming from the vicinity of $\varphi_+$ and $\varphi_-$. Moreover, $G_2^c$ does
not have the usual perturbative behavior $G_2^c=\frac{1}{2\mu^2}+\cO(g)$. This
is because there are contributions from two saddle points, which give a typical
nonperturbative result \cite{r7}. 

Let us compare the above result with the hypothetical $d>2$ case for which the
volume factor suppresses the tunneling (and so there is symmetry breaking).
Under this supposition we would consider contributions from only one of the two
vacua, $\varphi_+$ or $\varphi_-$. If we were to select $\varphi_-$, then $\frac
{Z[h]}{Z[0]}$ would be replaced by $\frac{Z^{(-)}[h]}{Z^{(-)}[0]}$. Thus, from
(\ref{e7}) we would find that
\begin{eqnarray}
\phantom{a}^{(-)}G_1^c&=&-\sqrt{\frac{6\mu^2}{g}}
+\frac{1}{4}\sqrt{\frac{3g}{2\mu^6}},\label{e8}\\
\phantom{a}^{(-)}G_2^c&=&\frac{1}{2\mu^2}+\frac{5g}{16\mu^6}.
\label{e9}
\end{eqnarray}
In (\ref{e8}) we observe the expected nonvanishing vacuum expectation value for
$\varphi$, which is given by the tree-level result $\varphi_-=-\sqrt{6\mu^2/g}$
plus a perturbative correction. Correspondingly, in (\ref{e9}) we see that the
propagator for the fluctuation around $\varphi_-$ has the typical perturbative
form $G_2^c=\frac{1}{2\mu^2}+\cO(g)$.

In the next section we will study the $\cPT$-symmetric version of the theory and
calculate the corresponding Green's functions. We will discover some intriguing
new properties, different from those of the conventional $\varphi^4$ theory. In
particular, we find that even though the formulas for the Green's functions in
(\ref{e8}) and (\ref{e9}) are only hypothetical, the correct formulas for the
Green's functions $G_1^c$ and $G_2^c$ of the corresponding $\cPT$-symmetric
theory can be obtained by analytic continuation of the formulas in (\ref{e8})
and (\ref{e9}). 

\section{$\cPT$-\,symmetric theory}
\label{s3}
In this section we study the behavior of the massive $\varphi^4$ theory in zero
dimensions and with negative coupling $-g$ ($g>0$). We examine this theory in
the presence of a linear source term $ih\varphi$ in the potential $V(\varphi)$: 
\begin{equation}
V(\varphi)=ih\varphi+\frac{m^2}{2}\varphi^2-\frac{g}{24}\varphi^4.
\label{e10}
\end{equation}
To this end, we calculate the partition function
\begin{eqnarray}
Z[h]=\sqrt{\frac{|m^2|}{2\pi}}\int_Cd\varphi\,e^{-V(\varphi)},
\label{e11}
\end{eqnarray}
where $C$ is a path of integration in the complex-$\varphi$ plane for which the
integral converges. In the present case this means that $C$ terminates inside of
two Stokes wedges of angular opening $\pi/4$, one centered about $-\pi/4$ and
the other centered about $-3\pi/4$ \cite{r1}. We also calculate the first two
connected Green's functions $G_1^c$ and $G_2^c$.

As in Sec.~\ref{s2}, we evaluate the integral in (\ref{e11}) in the
steepest-descent approximation, which consists of localizing the integrand
around a finite number of saddle points for which the real part of $-V$ is
maximal. On the steepest-descent contours ${\rm Im}\,V$ is constant. To locate
the saddle points we must solve the equation
\begin{eqnarray}
V'(\varphi)=ih+m^2\varphi-\frac{g}{6}\varphi^3=0.
\label{e12}
\end{eqnarray}
Then, we must find the two steepest-descent paths in the complex-$\varphi$
plane on which the saddle points are the maximum values of $-{\rm Re}\,V$.
These paths are also constant-phase contours (${\rm Im}\,V={\rm constant}$).
It is crucial to identify the constant-phase contour that can be deformed into
the original integration path $C$.

\subsection{The $m^2>m^2_c$ case}
\subsubsection{Saddle points}
Equation (\ref{e12}) has three solutions:
\begin{eqnarray}
\varphi_0&=&-i\beta,\label{e13}\\
\varphi_+&=&\frac{\sqrt{3}}{2}\alpha+i\frac{1}{2}\beta,\label{e14}\\
\varphi_-&=&-\frac{\sqrt{3}}{2}\alpha+i\frac{1}{2}\beta,\label{e15}
\end{eqnarray}
where $\alpha$ and $\beta$ are real numbers given by $\alpha=2m^2/A+A/g$ and
$\beta=2m^2/A-A/g$, and
$$A=\sqrt{-2gm^2_c}\,{\rm sgn}(B)|B|^{1/3}\quad{\rm and}\quad
B=-{\rm sgn}(h)+\sqrt{1-m^6/m_c^6}.$$
Since the theory is $\cPT$ symmetric, the solution $\varphi_+$, which has a
positive-real component, is transformed into the $\varphi_-$ solution, which has
a negative-real part, by a reflection through the imaginary axis. The solution
$\varphi_0$ lies on the imaginary axis. We expand (\ref{e13})--(\ref{e15}) for
small $h$ to obtain much simpler expressions for the saddle points:
$$\varphi_0=-i\frac{h}{m^2}+\cO(h^3),\quad \varphi_\pm=\pm\sqrt{\frac{6m^2}{g}}
+i\frac{h}{2m^2}\pm\frac{h^2}{8}\sqrt{\frac{3g}{2m^{10}}}+\cO(h^3).$$
Note that for $h>0$, $\varphi_+$ and $\varphi_-$ lie in the ${\rm Im}\,\varphi>
0$ half-plane and for $h<0$ they lie in the ${\rm Im}\,\varphi<0$ half-plane;
for positive values of $h$ the solution $\varphi_0$ lies on the
negative-imaginary axis and for negative values of $h$ it lies on the
positive-imaginary axis. For $h=0$ the saddle points all lie on the real axis.

To identify the steepest-descent curves, we let $\varphi=u+iv$ so that the real
part of the potential in (\ref{e10}) is
\begin{equation}
\label{e16}
{\rm Re}\,V(u,v)=\frac{m^2}{2}(u^2-v^2)-\frac{g}{24}(u^4-6u^2v^2+v^4)-hv.
\end{equation}
Then, from the saddle-point equation (\ref{e12}) we have 
$$\frac{\partial}{\partial u}{\rm Re}\,V=m^2u-\frac{g}{6}u^3+\frac{g}{2}uv^2=0,
\quad\frac{\partial}{\partial v}{\rm Re}\,V=-m^2v+\frac{g}{2}u^2v-\frac{g}{6}
v^3-h=0.$$

Next, we construct the Hessian matrix
$$H=\left(\begin{array}{cc}\frac{\partial^2}{\partial u\partial u}{\rm Re}\,V &
\frac{\partial^2}{\partial u\partial v}{\rm Re}\,V\\ \frac{\partial^2}{\partial
v\partial u}{\rm Re}\,V & \frac{\partial^2}{\partial v\partial v}{\rm Re}\,V
\end{array}\right)=\left(\begin{array}{cc} m^2-\frac{g}{2}(u^2-v^2) & guv\\
guv & -m^2+\frac{g}{2}(u^2-v^2)\end{array}\right),$$
and evaluate it at each of the saddle points (\ref{e13})--(\ref{e15}). The
eigenvectors associated with the positive eigenvalues of this matrix provide the
steepest-descent directions of $-V(u,v)$. On a steepest-descent path the
contribution is localized at the saddle point, so we can use Laplace's method to
evaluate the integral. For the saddle point $\varphi_0$, $u_0=0$ and $v_0=-
\beta$ [see (\ref{e13})] and thus
$$H(\varphi_0)=\left(\begin{array}{cc} m^2+\frac{g}{2}v_0^2 & 0\\ 0& -m^2
-\frac{g}{2}v_0^2\end{array}\right).$$
This matrix is already diagonal. One can easily see that for $m^2>m^2_c$ the
eigenvalue $\lambda_0=m^2+\frac{g}{2}v_0^2$ is positive, so the $u$ axis is the
direction of steepest descent from $\varphi_0$.

For the other two saddle points $\varphi_\pm$, $u_\pm=\pm\frac{\sqrt{3}}{2}
\alpha$ and $v_\pm=\frac{1}{2}\beta$ and thus
$$H(\varphi_\pm)=\left(\begin{array}{cc}-\frac{g}{3}u_\pm^2 & g u_\pm v_\pm\\
gu_\pm v_\pm & \frac{g}{3}u_\pm^2\end{array}\right).$$
For $u=u_+$ and $v=v_+$, the positive eigenvalue is 
\begin{equation}
\lambda_1^+=\frac{1}{4}g\alpha\sqrt{4\alpha^2+3\beta^2}
\label{e17}
\end{equation}
and the corresponding unit eigenvector is 
$$\vec{w}_1^+=\left(\frac{1}{\sqrt{2}}\frac{\sqrt{-2\alpha+\sqrt{4\alpha^2+3
\beta^2}}}{(4\alpha^2+3\beta^2)^{1/4}},{\rm sgn}(\beta)\frac{1}{\sqrt{2}}
\frac{\sqrt{2\alpha+\sqrt{4\alpha^2+3\beta^2}}}{(4\alpha^2+3\beta^2)^{1/4}}
\right).$$
The negative eigenvalue is
\begin{equation}
\lambda_2^+=-\frac{1}{4} g\alpha \sqrt{4\alpha^2+3\beta^2}
\label{e18}
\end{equation}
with unit eigenvector
$$\vec{w}^+_2=\left(-\frac{1}{\sqrt{2}}\frac{\sqrt{2\alpha+\sqrt{4\alpha^2+3
\beta^2}}}{(4\alpha^2+3\beta^2)^{1/4}},{\rm sgn}(\beta)\frac{1}{\sqrt{2}}\frac{
\sqrt{-2\alpha+\sqrt{4\alpha^2+3\beta^2}}}{(4\alpha^2+3 \beta^2)^{1/4}}
\right).$$

For $u=u_-$ and $v=v_-$ the eigenvalues $\lambda_1^-$ and $\lambda_2^-$ and the
corresponding unit eigenvectors $\vec{w}^-_1$ and $\vec{w}^-_2$ are obtained
from the previous ones by making the replacement $\alpha\to-\alpha$. An
inspection of (\ref{e17}) and (\ref{e18}) shows that for this case the
positive eigenvalue is $\lambda_1^-=-\lambda_2^+$ while the negative one is
$\lambda_2^-=-\lambda_1^+$. For notational simplicity, in the following we label
the two components of the unit vector $\vec{w}_1^+$ using the letters $a$ and
$b$: $\vec{w}_1^+\equiv(a,b)$. Also, we replace $\lambda_1^+$ ($\lambda_1^-$)
with $\lambda$. With this choice of parameters $a$ and $b$ we also get $\vec{
w}_2^+\equiv (-b,a)$, $\vec{w}_1^-\equiv(-a,b)$, and $\vec{w}_2^-\equiv(b,a)$.
As explained previously, for our calculation we are interested in the positive
eigenvalues and the corresponding eigenvectors. With the notation we have
introduced the positive eigenvalue is $\lambda$ in both cases ($\varphi_+$ and
$\varphi_-$) and the corresponding eigenvectors are $(a,b)$ and $(-a,b)$.

\subsubsection{Constant-phase contours}
We now determine the constant-phase contours in the $\varphi$ plane that are
also the steepest-descent paths from the saddle points. Our goal is to deform
the original path $C$ in (\ref{e11}) into a new set of steepest-descent contours
passing through the saddle points (\ref{e13})--(\ref{e15}). As explained
earlier, on this new path ${\rm Im}\,V$ is constant (or, more precisely,
piecewise constant), and the integral in (\ref{e11}) is approximated by a sum of
gaussian integrals localized at the saddle points.

The constant-phase contours are determined by the equation
\begin{equation}
\label{e19}
{\rm Im}\,V=hu+m^2uv-\frac{1}{6}gu^3v+\frac{1}{6}guv^3=c.
\end{equation}
The constant $c$ will take the values $c_+$ , $c_0$, or $c_-$ when we evaluate
(\ref{e19}) at $\varphi_+$, $\varphi_0$, or $\varphi_-$, respectively. We now
study the two cases $h>0$ and $h<0$ in turn. 

\begin{figure}[t!]
\vskip-4mm
\includegraphics[width=.60\textwidth]{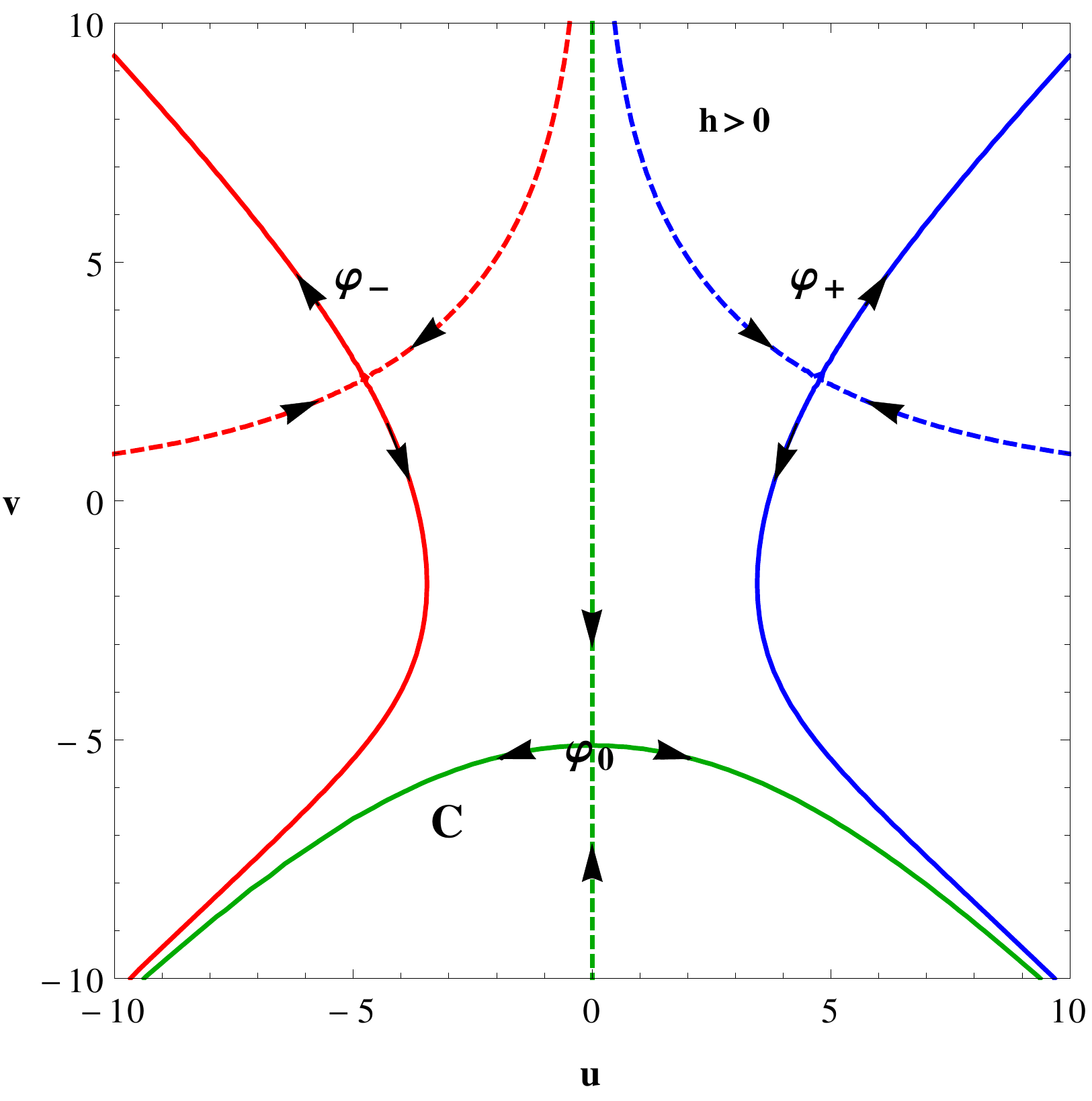}\vskip-4mm
\caption{[Color online] Constant-phase contours in the $(u,v)$ plane for $h>0$,
with the arbitrary choice $m^2=0.1$, $g=0.2$, and $h=5.0$. The constant-phase
contours associated with the saddle point $\varphi_0$ are green, those
associated with $\varphi_+$ are blue, and those associated with $\varphi_-$ are
red. The steepest-descent paths associated with these three saddle points are
solid, and the steepest-ascent paths emanating from the saddle points are
dashed. The arrows indicate the {\it down} directions. The original contour
terminates in Stokes wedges in the southwest and southeast quadrants of the
complex plane. Therefore, for the case $h>0$ the only relevant saddle point is
$\varphi_0$, and the original integration path must be deformed into the contour
labeled $C$. For this case the saddle points $\varphi_\pm$ play no role in the
asymptotic evaluation of the partition function.}
\label{F1}
\end{figure}

\vspace{.2cm}
\noindent{\it The $h>0$ case:} We first consider the case $h>0$. Figure~\ref{F1}
shows the saddle points $\varphi_0$, $\varphi_+$, and $\varphi_-$ of
(\ref{e13})--(\ref{e15}) in the $(u,v)$ plane for a specific (although
irrelevant) numerical choice of the parameters together with the constant-phase
contours passing through these points. The solid lines are the steepest-descent
(constant-phase) contours associated with the three saddle points. The phases
are $c_+$, $c_-$, and $c_0$, respectively. 

In general, to evaluate the partition function $Z[h]$ of (\ref{e11}) in the
saddle-point approximation, we must identify the contour $C$ that passes through
one or more of the saddle points and that at the same time is a steepest-descent
path that terminates inside the Stokes wedges in the southwest and southeast
quadrants of the complex plane. An inspection of Fig.~\ref{F1} shows that when
$h>0$, of the three saddle points only one of them, namely $\varphi_0$, fulfills
these conditions and that the solid curve labeled $C$ that passes through
$\varphi_0$ is the desired contour. For $\varphi=\varphi_0=iv_0$ we have ${\rm
Im}\,V(\varphi_0)=0$ [see (\ref{e19})], so the constant-phase contour $C$ is
given by (\ref{e19}) with $c=0$; thus, the equation for the curve $C$ is
$$\frac{6h}{gv}+\frac{6m^2}{g}+v^2-u^2=0.$$
This curve terminates in the middle of the Stokes wedges. By evaluating the
integral in (\ref{e11}) along this constant-phase contour $C$, we obtain
$$Z[h]=\sqrt{\frac{|m^2|}{2\pi}}\int_Cd\varphi\,e^{-V(\varphi)}=\sqrt{\frac{
|m^2|}{2\pi}}e^{-i c_0}\int^\infty_{-\infty}du\,\left(1+i\frac{d v(u)}{du}
\right)e^{-{\rm Re}\,V[u,v(u)]}.$$

Expanding the real part of the potential around the saddle point ($u_0=0$,
$v_0=-\beta$) to second order in $(u-u_0)$ and $(v-v_0)$, we get
[with ${\rm Re}\,V(\varphi_0)={\rm Re}\,V_0$]
$${\rm Re}\,V[u,v(u)]\sim{\rm Re}\,V_0+\frac{1}{2}\left(u-u_0,v(u)-v_0\right)
\cdot H(\varphi_0)\cdot\left(\begin{array}{c} u-u_0\\ v(u)-v_0\end{array}
\right).$$
Therefore, the saddle point approximation for $Z[h]$ is
\begin{equation}
\label{e20}
Z[h]=\exp\left(-{\rm Re}\,V_0-\frac{1}{2}\ln\frac{\lambda_0}{m^2}\right),
\end{equation}
where ${\rm Re}\,V_0=-m^2\beta^2/2-g\beta^4/24+h\beta$ and $\lambda_0=m^2+g
\beta^2/2$.

We expand the partition function $Z[h]$ in (\ref{e20}) in powers of $h$ 
($\beta\sim\frac{h}{m^2}$) and get
\begin{equation}
\label{e21}
Z[h]=1-\left(\frac{1}{m^2}+\frac{g}{2m^6}\right)\frac{h^2}{2}+\cO(h^4).
\end{equation}
From this result we can obtain the one-point and two-point Green's functions
$G_1=i\frac{1}{Z[0]}\left.\frac{d Z[h]}{dh}\right|_{h=0}$ and
$G_2=(i)^2\frac{1}{Z[0]}\left.\frac{d^2 Z[h]}{dh^2}\right|_{h=0}$.
Finally, the connected Green's functions are
\begin{eqnarray}
G_1^c&=&0,\label{e22}\\ 
G_2^c&=&\frac{1}{m^2}+\frac{g}{2m^6}.\label{e23}
\end{eqnarray}

We emphasize that despite the negative sign of the quartic coupling constant
($-g<0$) (which one might interpret as implying instability due to unboundedness
below), the function $G_2$ is real and positive as a consequence of $\cPT$
symmetry. It is also worth noting that in this case the partition function
(\ref{e20}) and the two-point Green's function (\ref{e23}) has the form one
would expect from a perturbative expansion in powers of the coupling constant
$g$. In other words, the saddle-point approximation to $Z[h]$ gives, as in the
conventional positive-coupling constant theory, the usual perturbative
expansion. For this reason, we call the saddle point $\varphi_0$ the {\it
perturbative} saddle point.

Let us compare the partition function (\ref{e21}) of the $\cPT$-symmetric theory
with the corresponding partition function (\ref{e5}) of the conventional theory
for the $m^2>m^2_c$ case. Note first that the potential (\ref{e10}) of the
$\cPT$-symmetric theory is obtained from the potential (\ref{e2}) of the
conventional theory by making the replacements $g\to-g$ and $h\to ih$. By making
the same replacements in the partition function (\ref{e5}) of the ordinary
theory, we obtain the partition function (\ref{e21}) of the $\cPT$-symmetric
theory. Thus, by analytically continuing the partition function of the
conventional theory to imaginary values of $h$ and negative values of $g$, the
partition function of the corresponding $\cPT$-symmetric theory is obtained. We
will see in the following that this simple connection between the two theories
is lost when the partition function of one of the two theories gets
contributions from more than one saddle point while the partition function of
the other theory gets a contribution from only one saddle point.

\begin{figure}[t]
\vskip-4mm
\includegraphics[width=.60\textwidth]{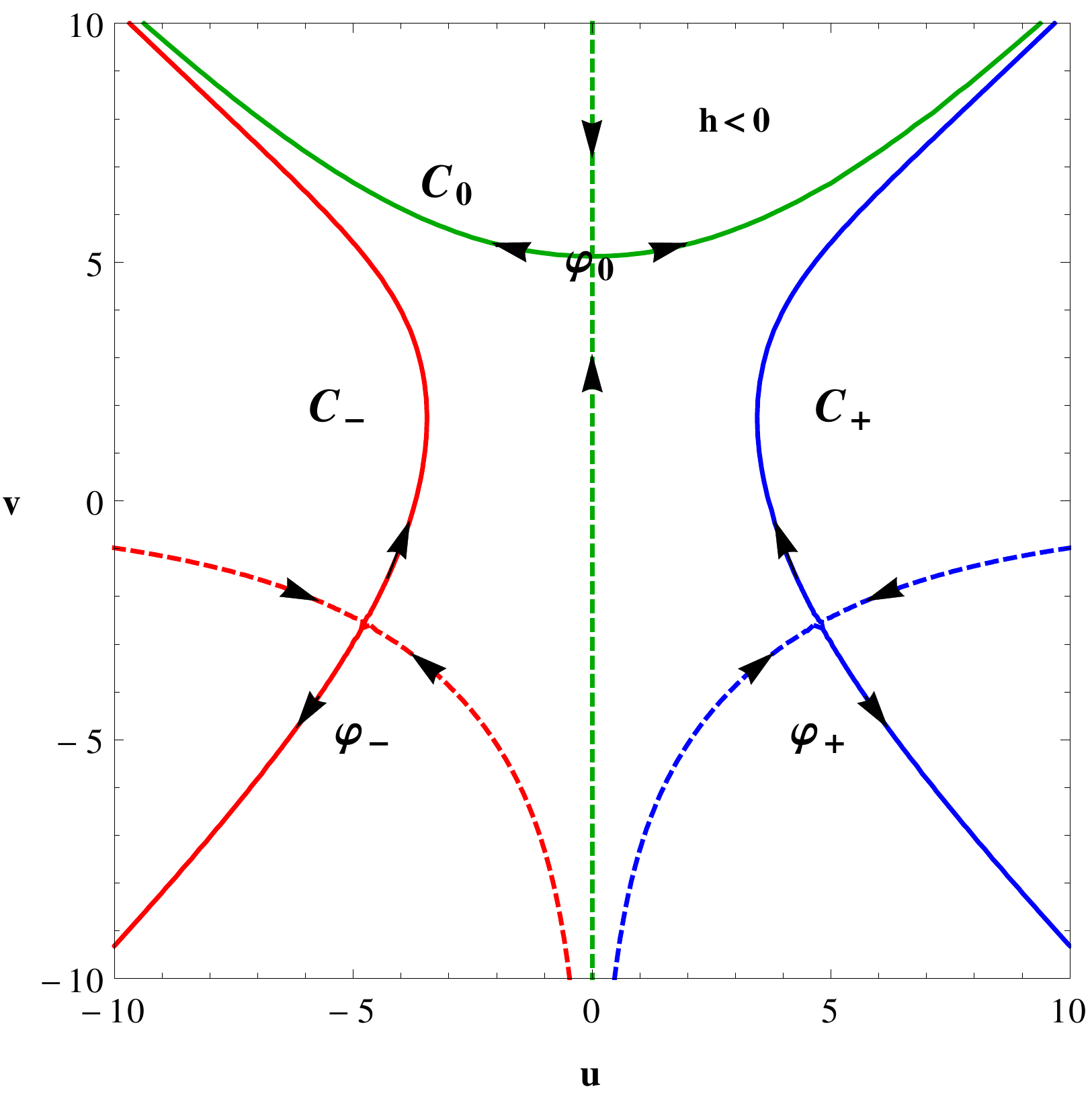}\vskip-4mm
\caption{[Color online] Constant-phase contours in the $(u,v)$ plane for the
case $h<0$ with the arbitrary choice $m^2=0.1$, $g=0.2$, and $h=-5.0$. As in
Fig.~\ref{F1}, the constant-phase contours associated with the saddle points
$\varphi_0$, $\varphi_+$, and $\varphi_-$ are colored green, blue, and red. The
steepest-descent paths associated with these three saddle points are solid, and
the steepest-ascent paths emanating from the saddle points are dashed and the
arrows indicate the {\it down} directions. As in the case $h>0$, the original
integration contour terminates in Stokes wedges in the southwest and southeast
quadrants of the complex plane. However, unlike Fig.~\ref{F1}, all three saddle
points are relevant. The original integration path must be deformed into three
constant-phase contours labeled $C_-$, $C_0$, and $C_+$, which are joined
end-to-end.}
\label{F2}
\end{figure}

\vspace{.2cm}
\noindent{\it The $h<0$ case}: When $h<0$, the saddle points $\varphi_0$,
$\varphi_+$, and $\varphi_-$ and the constant-phase curves are obtained from
those of the $h>0$ case by reflecting about the $v=0$ axis (compare
Figs.~\ref{F2} and \ref{F1}). This has crucial consequences for the saddle-point
approximation of $Z[h]$ in (\ref{e11}). Indeed, Fig.~\ref{F2} shows that the
contour needed to evaluate the integral is quite different from the contour $C$
of Fig.~\ref{F1}. As shown in Fig.~\ref{F2}, the integration path is now
obtained by joining three constant-phase contours, $C_-$, $C_0$, and $C_+$.

For $h<0$ (unlike the $h>0$ case) all three saddle points contribute to the
integral in (\ref{e11}). Thus, along the path $C_-+C_0+C_+$ we decompose the
integral as follows:
\begin{equation}
Z[h]=\sqrt{\frac{|m^2|}{2\pi}}\left(\int_{C_-}\,d\varphi \,e^{-V(\varphi)}+
\int_{C_0}\,d\varphi\,e^{-V(\varphi)}+\int_{C_+}\,d\varphi\,e^{-V(\varphi)}
\right).
\label{e24}
\end{equation}
We consider each of these integrals in turn: For the integral along $C_0$, the
calculation is the same as for the $h>0$ case, and we obtain
\begin{equation}
\label{e25}
\sqrt{\frac{|m^2|}{2\pi}}\int_{C_0}\,d\varphi\,e^{-V(\varphi)}\sim\exp
\left(-{\rm Re}\,V_0-\frac{1}{2}\ln\frac{\lambda_0}{m^2}\right).
\end{equation}

The curve $C_+$ is given by (\ref{e19}), and using (\ref{e14}) we get
$$c\equiv c_+={\rm Im}\,V(\varphi_+)=h\frac{\sqrt{3}}{2}\alpha+\frac{\sqrt{3}}
{4}m^2\alpha\beta -\frac{\sqrt{3}}{32}g\alpha^3\beta+\frac{g}{32\sqrt{3}}\alpha
\beta^3.$$
We parametrize $C_+$ in terms of the variable $t$, $u=u_+(t)$ and $v=v_+(t)$ and
note that the vector $\vec{w}_1^+=(a,b)$ is the unit vector tangent to $C_+$ in
$\varphi_+=u_++iv_+=\frac{\sqrt{3}}{2}\alpha+i\frac{1}{2}\beta$ [see
(\ref{e14})]. Expanding $u_+(t)$ and $v_+(t)$ to first order in $t$, we get 
$u_+(t)-u_+\sim at$, $v_+(t)-v_+\sim bt$, so the expansion of ${\rm Re}\,
V(\varphi)$ around $\varphi_+$ to second order in $t$ is
\begin{eqnarray}
{\rm Re}\,V(\varphi)&=&{\rm Re}\,V(\varphi_+)+\frac{1}{2}\left[u_+(t)-u_+,v_+(t)
-v_+\right]\cdot H(\varphi_+)\cdot\left(\begin{array}{c}u_+(t)-u_+\\ v_+(t)-v_+
\end{array}\right)\nonumber\\
&=&{\rm Re}\,V(\varphi_+)+\frac{1}{2}\lambda t^2,\nonumber
\end{eqnarray}
where we have used $a^2+b^2=1$. The integral along $C_+$ is then approximated by
\begin{eqnarray}
\label{e26}
\sqrt{\frac{|m^2|}{2\pi}}\int_{C_+}d\varphi\,e^{-V(\varphi)}\sim(a+ib)\,\exp
\left[-{\rm Re}\,V(\varphi_+)-\frac{1}{2}\ln\frac{\lambda}{m^2}-ic_+\right].
\end{eqnarray}

Last, we calculate the contribution to $Z[h]$ in (\ref{e24}) coming from the
integral along the curve $C_-$. Inserting (\ref{e15}) into (\ref{e16}) and
(\ref{e19}), we find that $c_-=-c_+$ and ${\rm Re}\,V(\varphi_-)={\rm Re}\,V(
\varphi_+)$. Around $\varphi_-$, the curve $C_-$ is approximated as $u_-(t)-
u_-\sim-at$ and $v_-(t)-v_-\sim bt$. Following the same steps as for the
evaluation of the integral along $C_+$, we find that
\begin{eqnarray}
\label{e27}
\sqrt{\frac{|m^2|}{2\pi}}\int_{C_-}d\varphi\,e^{-V(\varphi)}\sim(a-ib)\exp\left[
-{\rm Re}\,V(\varphi_-)-\frac{1}{2}\ln \frac{\lambda}{m^2}-ic_-\right].
\end{eqnarray}
Combining (\ref{e25}), (\ref{e26}), and (\ref{e27}), and using the notation
$\overline c\equiv c_+=-c_-$ and ${\rm Re}\,\overline V\equiv{\rm Re}\,V(
\varphi_+)={\rm Re}\,V(\varphi_-)$, the final result for the saddle-point
approximation to $Z[h]$ is
\begin{equation}
\label{e28}
Z[h]=\exp\left(-{\rm Re}\,V_0-\frac{1}{2}\ln\frac{\lambda_0}{m^2}\right)+2
\left(a\cos \overline c+b\sin \overline c\right)
\exp\left(-{\rm Re}\,\overline V-\frac{1}{2}\ln\frac{\lambda}{m^2}\right).
\end{equation}

The contribution to $Z[h]$ for $h<0$ that appears in the first term of
(\ref{e28}) agrees with the result for $Z[h]$ in (\ref{e20}) for $h>0$.
However, in (\ref{e28}) an additional term appears, which comes from the
contribution of the two additional nontrivial saddle points $\varphi_+$ and
$\varphi_-$. These saddle points {\it do not} contribute to $Z[h]$ in the $h>0$
case. We emphasize that the $h>0$ and the $h<0$ cases are different because the
saddle-point contour passes only through the $\varphi_0$ saddle point when $h>0$
(see Fig.~\ref{F1}), while the saddle-point contour passes through all three
saddle points, $\varphi_-$, $\varphi_0$, $\varphi_+$, when $h<0$ (see 
Fig.~\ref{F2}). Thus, the $h>0$ and the $h<0$ cases cannot be obtained from one
another by simply changing the sign of $h$ in the partition function.
 
The peculiar discontinuity in the partition functions for the $h>0$ case
(\ref{e20}) and the $h<0$ case (\ref{e28}) only occurs in the $\cPT$ symmetric
theory and does not appear in the conventional $\varphi^4$ theory. As a
consequence, the Green's functions obtained from (\ref{e20}), which are the
limit as $h\to0^+$ of derivatives of $Z[h]$, are different from the Green's
functions obtained from (\ref{e28}), which appear in the limit $h\to0^-$. To see
this difference explicitly, we expand (\ref{e28}) in powers of $h$. Keeping only
terms up to $\cO(h^2)$ we get
$$\frac{Z[h]}{Z[0]}=1+\left(\sqrt{\frac{6 m^2}{g}}+\frac{1}{8m^2}\sqrt{\frac{3g}
{2m^2}}\right)e^{-\frac{3m^4}{2g}}h-\left(\frac{1}{m^2}+\frac{g}{2m^6}\right)
\frac{h^2}{2},$$
from which we read off the connected Green's functions $G_1^c$ and $G_2^c$:
\begin{eqnarray}
G_1^c &=& i\left(\sqrt{\frac{6m^2}{g}}+\frac{1}{8m^2}
\sqrt{\frac{3g}{2m^2}}\right)e^{-\frac{3m^4}{2g}},\label{e29}\\
G_2^c&=&\frac{1}{m^2}+\frac{g}{2m^6}.\label{e30}
\end{eqnarray}
Note that the two-point Green's functions $G_2^c$ in (\ref{e23}) and (\ref{e30})
for the cases $h>0$ and $h<0$ are the same. However, the one-point Green's
functions $G_1^c$ in (\ref{e22}) and (\ref{e29}) are different. This difference
is a violation of the usual Bogoliubov limit \cite{r8} for constructing the
Green's functions. For the conventional theory, the Green's functions are
obtained as the $h\to0$ limit of $Z[h]$ derivatives, independently of the sign
of $h$.

\subsection{The $m^2<m^2_c$ case}
Let us consider now the evaluation of $Z[h]$ in (\ref{e11}) for the case $m^2=-
\mu^2<m^2_c$ (where $\mu^2>0$). As before, we first look for the saddle points
and then determine the constant-phase contours for the saddle-point
approximation of $Z[h]$.

\subsubsection{Saddle points}
The saddle points are given by the solutions to (\ref{e12}) and are purely
imaginary:
\begin{equation}
\label{e31}
\varphi_\pm=i\rho\cos\left(\frac{|\theta|+2\pi\Theta(\pm h)}{3}\right),\quad
\varphi_0=i\rho\cos\left(\frac{|\theta|-2\pi}{3}\right),\quad
\end{equation}
where 
$$\rho=-2\,{\rm sgn}(h)\sqrt{\frac{2\mu^2}{g}},\quad\theta=-{\rm sgn}(h)\,
\mbox{arctan}\left(\sqrt{-1+\frac{8\mu^6}{9gh^2}}\right).$$
For any $h$, $\varphi_-$ lies on the negative-imaginary axis in the
complex-$\varphi$ plane, $\varphi_+$ lies on the positive-imaginary axis, and
depending on the sign of $h$, $\varphi_0$ lies on the positive-imaginary or
negative-imaginary axis. Expanding the saddle-point solutions in (\ref{e31}) 
in $h$, we get 
$$\varphi_0=i\frac{h}{\mu^2}+\cO(h^3),\quad\varphi_\pm=i\left(\pm\sqrt{\frac{6
\mu^2}{g}}-\frac{h}{2\mu^2}\mp\sqrt{\frac{3g}{2\mu^{10}}}\frac{h^2}{8}\right)+
\cO(h^3).$$

Next, letting $\varphi=u+iv$, we find that the real part of the potential
(\ref{e10}) is
$${\rm Re}\,V(u,v)=-\frac{\mu^2}{2}(u^2-v^2)-\frac{g}{24}(u^4-6u^2v^2+v^4)-hv.$$
We then calculate the Hessian matrix, which at the three saddle points
(\ref{e31}) is diagonal:
\begin{equation}
\label{e32}
H=\left(\begin{array}{cc}\frac{\partial^2}{\partial u\partial u}{\rm Re}\,V &
\frac{\partial^2}{\partial u\partial v}{\rm Re}\,V \\ \frac{\partial^2}{\partial
v\partial u}{\rm Re}\,V & \frac{\partial^2}{\partial v\partial v}{\rm Re}\,V
\end{array}\right)=\left(\begin{array}{cc} -\mu^2+\frac{g}{2}v^2 & 0 \\
0 & \mu^2-\frac{g}{2}v^2\end{array}\right),
\end{equation}
where $v$ is $v_0=i\varphi_0$ or $v_\pm=\varphi_\pm$, depending on which of the
saddle points we consider.

As before, the local directions of the steepest-descent paths from each of the
saddle points are determined by finding the eigenvectors of the Hessian matrix
with $u$ and $v$ at the saddle points. Because $H$ has a diagonal form, the
eigendirections are parallel to the $u$ and the $v$ axes. The signs of the two
eigenvalues $-\mu^2 +\frac{g}{2}v^2$ and $\mu^2-\frac{g}{2}v^2$ are found by
substituting $v_+$, $v_-$, and $v_0$ [see (\ref{e31})] in (\ref{e32}). We find
that $-\mu^2+\frac{g}{2}v^2$ is negative for $\varphi_0$ and positive for
$\varphi_+$ and $\varphi_-$. The opposite is true for $\mu^2-\frac{g}{2}v^2$.

\subsubsection{Constant-phase contours}
\begin{figure}[t]
\vskip-4mm
\begin{minipage}{0.49\textwidth}
\includegraphics[width=.95\textwidth]{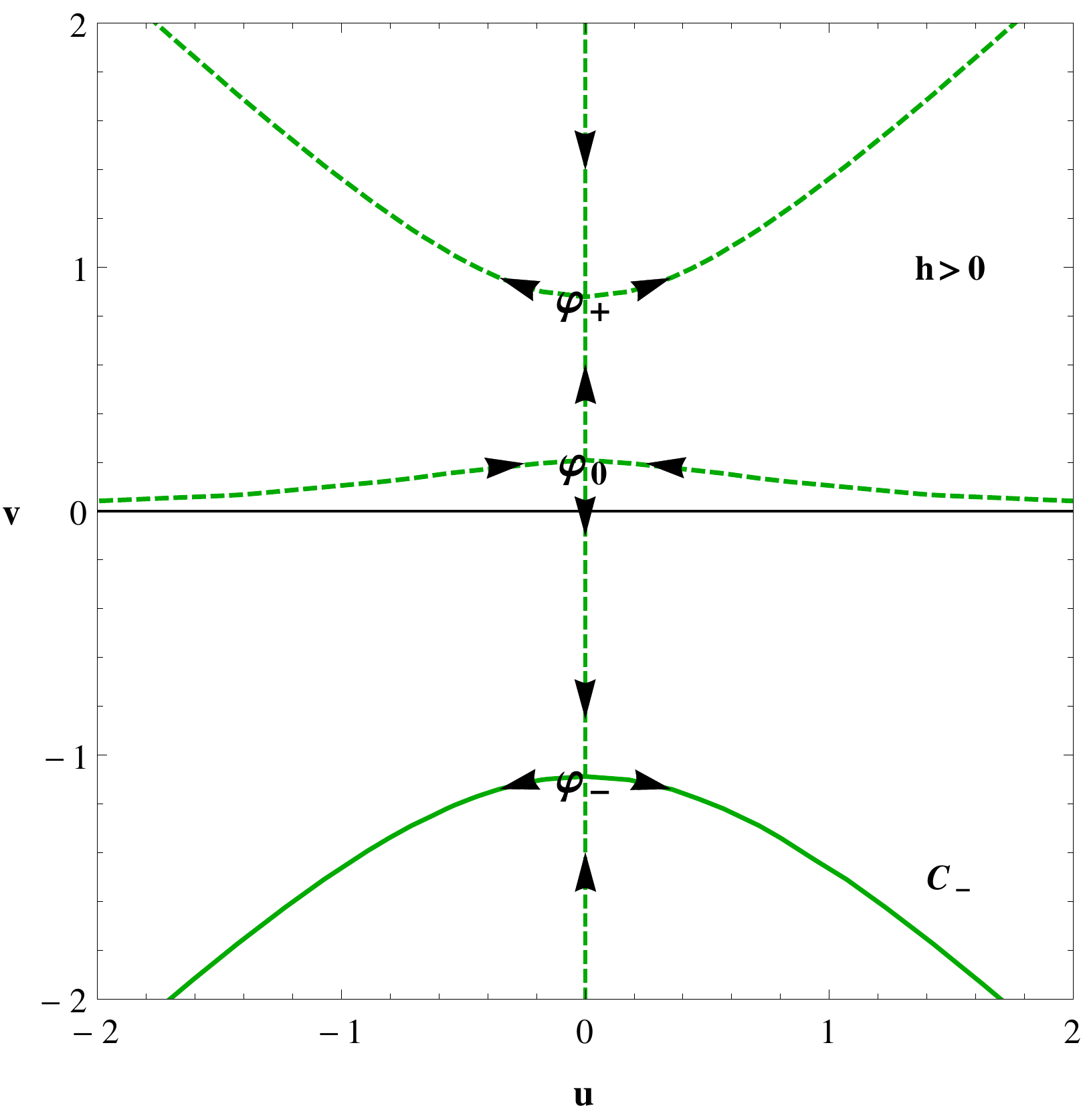}
\end{minipage}
\begin{minipage}{0.49\textwidth}
\includegraphics[width=.95\textwidth]{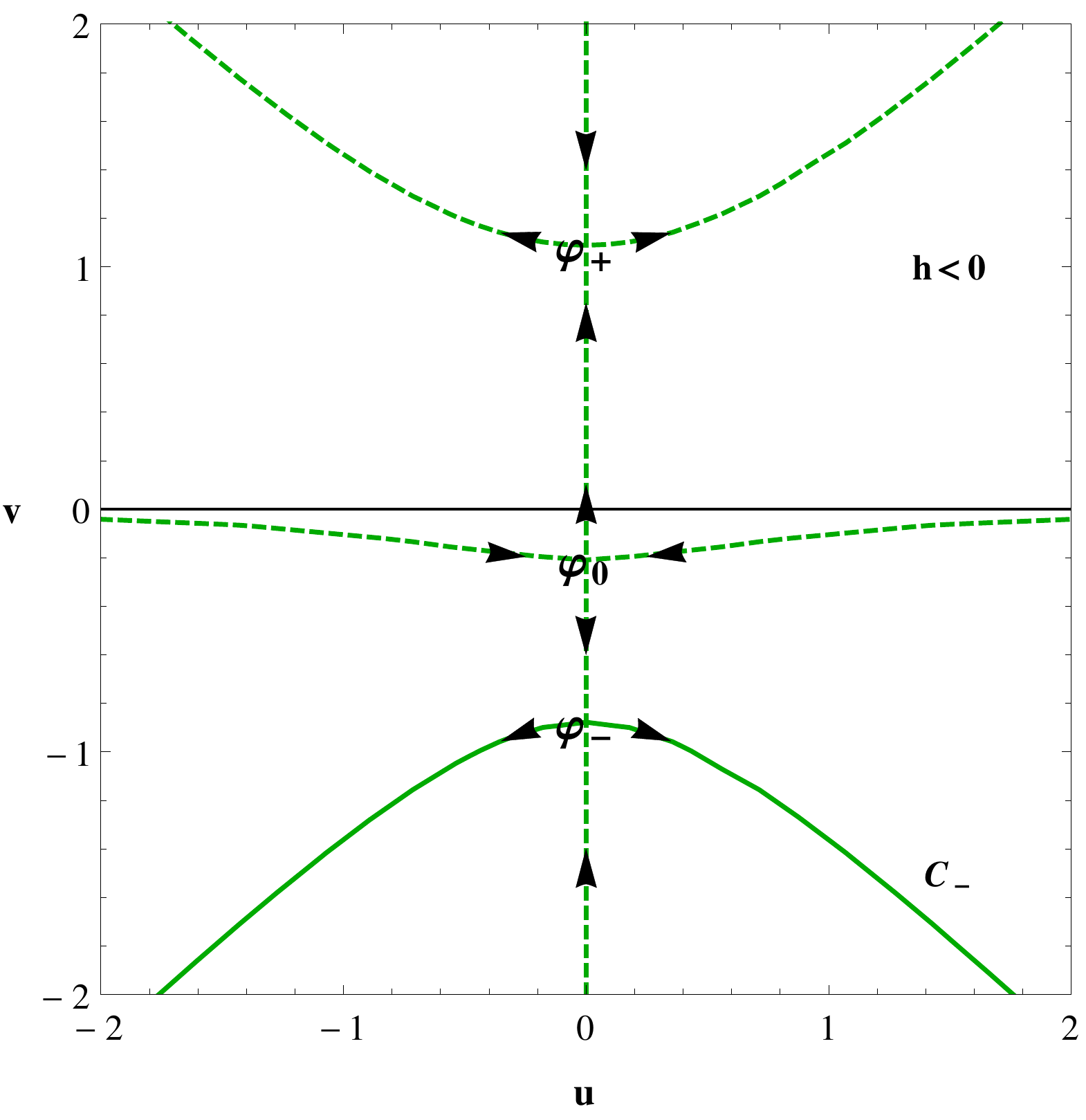}
\end{minipage}
\vskip-4mm
\caption{Constant-phase contours in the $(u,v)$ plane for the potential $V$ in
(\ref{e10}) for the parameters $m^2=-0.1$, $g=0.6$. The left panel portrays the
case $h>0$ for the specific value $h=0.02$ and the right panel portrays the case
$h<0$ for the specific value $h=-0.02$. One contour lies on the imaginary axis
and the other three contours emerge horizontally from each of the three saddle
points $\varphi_+$, $\varphi_-$, and $\varphi_0$. The only relevant contour is
the one that passes through $\varphi_-$ because this contour terminates in the
Stokes wedges, which lie in the southwest and southeast quadrants of the $(u,v)$
plane. This contour is labeled $C_-$ and is represented as a solid line. All
other contours are indicated by dashed lines.}
\label{F3}
\end{figure}

To find the constant-phase contours we must solve the equation
\begin{equation}
\label{e33}
{\rm Im}\,V=hu-\mu^2uv-gu^3v/6+guv^3/6=c,
\end{equation}
where $V$ is the potential in (\ref{e10}) and the constant $c$ is determined by
evaluating ${\rm Im}\,V$ at each of the saddle points $\varphi_+$, $\varphi_0$,
and $\varphi_-$. The real part of all three saddle points vanishes, so from
(\ref{e33}), we have $c=0$ for each saddle point. Factoring (\ref{e33}) gives a
linear equation, namely $u=0$, and a cubic equation for the constant-phase
contours:
\begin{equation}
v^3-\left(6\mu^2/g+u^2\right)v+6h/g=0.
\label{e34}
\end{equation} 
The $u=0$ contour is just the $v$ axis and passes through all three saddle
points $\varphi_+$, $\varphi_-$, and $\varphi_0$. The cubic equation (\ref{e34})
provides the remaining three constant-phase contours, one for each of the three
saddle points. The constant-phase contours are plotted in Fig.~\ref{F3} for a
specific choice of the parameters. In the left (right) panel we consider the $h>
0$ ($h<0$) case. Observe that in both cases, the only path that terminates in
the Stokes wedges is the path that passes through $\varphi_-$. This path is
labeled $C_-$.

Laplace's method along the path $C_-$ approximates $Z[h]$ as $\exp\left(
-{\rm Re}\,V_--\frac{1}{2}\ln\frac{\lambda_-}{m^2}\right)$, where ${\rm Re}\,V_-
\equiv{\rm Re}\,V(\varphi_-)=\mu^2v_-^2/2-gv_-^4/24-h v_-$ and $\lambda_-=-\mu^2
+gv_-^2/2>0$. Thus, the normalized partition function expanded in powers of $h$ is
$$\frac{Z[h]}{Z[0]}=1-\left(\sqrt{\frac{6\mu^2}{g}}+\frac{1}{4}\sqrt{\frac{3g}{
2\mu^6}}\right)h+\left(\frac{1}{\mu^2}+\frac{6\mu^2}{g}+\frac{13g}{32\mu^6}
\right)\frac{h^2}{2}+\cO(h^3).$$
This equation determines the first two connected Green's functions:
\begin{equation}
G_1^c=-i\left(\sqrt{\frac{6\mu^2}{g}}+\frac{1}{4}\sqrt{\frac{3g}{2\mu^6}}
\right),\quad G_2^c=\frac{1}{2\mu^2}-\frac{5}{16}\frac{g}{\mu^6}.
\label{e35}
\end{equation}
Note that the connected two-point Green's function $G_2^c$ is precisely what
one obtains from perturbation theory. In particular, the squared mass of the
fluctuation above the vacuum is $\mu^2=-2m^2$, as in the conventional theory.
In contrast, $G_1$ is nonvanishing. Evidently, the $\cPT$-symmetric theory
evades the MWC theorem \cite{r9} (according to which there cannot be spontaneous
symmetry breaking below $d=2$ dimensions).

\section{Summary and Conclusions}
\label{s4}
We have studied the zero-dimensional Euclidean $\cPT$-symmetric $\varphi^4$
theory in the presence of an external source $h$ and have compared it with the
corresponding conventional $\varphi^4$ theory. We calculated the partition
function and the Green's functions using the saddle-point approximation. As is
the case for the conventional theory, the $\cPT$-symmetric theory possess
distinct phases depending on whether the squared mass $m^2$ is greater than or
less than the critical value $m^2_c=-(g h^2)^{1/3}$. These phases are
characterized by different Green's functions. Furthermore, when $m^2>m^2_c$, the
$\cPT$-symmetric theory exhibits two distinct subphases depending on whether
$h\to0^+$ or $h\to0^-$.

The subphases can be described as follows: For $m^2>m^2_c$ and for $h>0$, the
partition function and the Green's functions of the $\cPT$-symmetric theory can
be obtained from the corresponding functions of the conventional theory by
analytic continuation to imaginary values of the external source $h$ ($h\to+ih$)
and to negative values of the coupling $g$ ($g\to-g$). This connection between
the two theories arises because in both cases the partition function receives
contributions from one saddle point only; the saddle point of the
$\cPT$-symmetric theory can be obtained from the saddle point of the
conventional theory by an anticlockwise rotation of $\pi/2$ in the
complex-$\varphi$ plane. On the other hand, for $m^2>m^2_c$ and $h<0$, while the
partition function of the conventional theory is still dominated by the
perturbative saddle point, the partition function of the $\cPT$-symmetric theory
receives contributions from the rotated perturbative saddle point and {\it also
from two nonperturbative saddle points}. This subphase behavior of the
$\cPT$-symmetric theory is a new and unexpected feature. The Green's functions
obtained by performing the two limits $h\to0^+$ and $h\to0^-$ are different, and
this is a clear violation of the usual Bogoliubov calculation of the Green's
functions.

The $\cPT$-symmetric theory differs from the conventional theory in other
respects. In addition to the behavior of the $m^2>m^2_c$ phase, new phenomena
appear in the $m^2<m^2_c$ phase. Spontaneous symmetry breaking in the
conventional quartic theory is forbidden for $d<2$ due to tunneling
between the degenerate minima; this absence of symmetry breaking is due to the
MWC theorem \cite{r9}. The MWC theorem certainly holds for $d=0$ and it implies
that for the conventional theory the one-point Green's function $G_1^c$ vanishes
even when $m^2<m^2_c$. In the context of the saddle-point approximation, the
vanishing of $G_1^c$ is due to the cancellation of contributions to the
partition function coming from the two nonperturbative saddle points. Thus, the
$\cPT$-symmetric theory evades the MWC theorem in much the same way that
supersymmetry evades the Coleman-Mandula theorem, namely, by circumventing the
assumptions needed to prove it.

However, for the $\cPT$-symmetric quartic theory the dominant contribution to
$Z[h]$ comes from just one saddle point. Thus, the $m^2<m^2_c$ phase of the
$\cPT$-symmetric theory is characterized by a nonvanishing $G_1$ when $d=0$.
Hence, we would expect that even when $0<d<2$, in the $\cPT$-symmetric
$\varphi^4$ theory a phase transition will occur that is not triggered by the
usual spontaneous-symmetry-breaking mechanism. Moreover, we have demonstrated
the surprising result that while $G_1=0$ in the conventional theory, $G_1\neq0$
when $m^2>m^2_c$ in the $\cPT$-symmetric theory.

Let us make a further comparison between the conventional and the 
$\cPT$-symmetric quartic theories. We have seen that when $m^2>m^2_c$ and $h<0$,
the partition function of the $\cPT$-symmetric theory is not an analytic
continuation of the corresponding partition function of the conventional theory.
This is because the $\cPT$-symmetric $Z[h]$ receives contributions from three
saddle points, while only one saddle point contributes to the partition function
of the conventional theory. In the $m^2<m^2_c$ case, the situation is similar
although the roles of the ordinary and the $\cPT$-symmetric theories are
exchanged. The partition function of the conventional theory receives
contributions from two saddle points, while $Z[h]$ of the corresponding
$\cPT$-symmetric theory gets contributions from only the one saddle point
$\varphi_-$ (see Fig.~\ref{F3}). We stress that a cancellation leading to
$G_1=0$ occurs when there are two saddle points, but this is impossible when
there is only one saddle point.

We have pointed out in Sec.~\ref{s2} in our discussion of the conventional
theory that we can artificially induce spontaneous symmetry breaking by
discarding the contribution to $Z[h]$ from $\varphi_+$. For $d<2$ this procedure
is artificial because both the left and right saddle points contribute to $Z[h]$
so the theory cannot exhibit spontaneous symmetry breaking. Of course, one of
the two saddle points must be discarded only when $d>2$. We have entertained
this possibility for the $d=0$ conventional theory in Sec.~\ref{s2} only to
compare the Green's functions in (\ref{e35}) with the analytic continuation (a
rotation in the complex-$\varphi$ plane) of the Green's functions in (\ref{e8})
and (\ref{e9}). 

To conclude, the structure of the quartic $\cPT$-symmetric theory in zero
dimensions is much richer than that of conventional quartic theory. It is also
far more elaborate than that of the cubic $i\varphi^3$ theory \cite{r2,r3}. Our
immediate objective now is to extend the results for the $\cPT$-symmetric
quartic theory to the $d>0$ case, just as we did for the cubic theory.

\acknowledgments
CMB thanks the Royal Society (U.K.) for a travel grant and VB thanks the
Istituto Nazionale di Fisica Nucleare (INFN) for financial support.

\end{document}